\documentclass[iop,apj]{emulateapj}
\usepackage{float}
\usepackage{amsmath}
\usepackage{rotating}
\usepackage{amssymb}
\usepackage{hyperref}

\def\gtrsim{\mathrel{\hbox{\rlap{\hbox{\lower4pt\hbox{$\sim$}}}\hbox{$>$}}}}
\def\lesssim{\mathrel{\hbox{\rlap{\hbox{\lower4pt\hbox{$\sim$}}}\hbox{$<$}}}}
\def\gtrsim{\mathrel{\hbox{\rlap{\hbox{\lower4pt\hbox{$\sim$}}}\hbox{$>$}}}}
\def\farcs{\hbox{$.\!\!^{\prime\prime}$}}
\def\farcm{\hbox{$.\!\!^{\prime}$}}

\begin{document}

\def\chan{{\sl Chandra\ }}

\title{Deep Chandra Observations of the Pulsar Wind Nebula Created by PSR B0355+54}

\author{Noel Klingler$^1$, Blagoy Rangelov$^1$, Oleg Kargaltsev$^1$, George G.~Pavlov$^2$, Roger W.~Romani$^3$, Bettina Posselt$^2$, Patrick Slane$^4$, Tea Temim$^{5,6}$, C.-Y.~Ng$^7$, Niccol\`o Bucciantini$^{8,9}$, Andrei Bykov$^{10,11,12}$, Douglas A.~Swartz$^{13}$, Rolf Buehler$^{14}$}
\affil{$^1$The George Washington University, Department of Physics, 725 21st Street NW, Washington, DC 20052, USA \\ 
$^2$Pennsylvania State University, Department of Astronomy \& Astrophysics, 525 Davey Laboratory, University Park, PA 16802, USA \\ 
$^3$Stanford University, Department of Physics, 382 Via Pueblo, Stanford, CA 94305, USA \\
$^4$Harvard-Smithsonian Center for Astrophysics, 60 Garden Street, Cambridge, MA 02138, USA \\
$^5$Observational Cosmology Lab, Code 665, NASA Goddard Space Flight Center, Greenbelt, MD 20771, USA \\
$^6$CRESST, University of Maryland -- College Park, College Park, MD 20742, USA \\
$^7$Department of Physics, The University of Hong Kong, Pokfulam Road, Hong Kong \\
$^8$INAF -- Osservatorio Astrofisico di Arcetri, L.go E.~Fermi 5, I-50125 Firenze, Italy \\
$^{9}$INFN -- Sezione di Firenze, Via G.~Sansone 1, I-50019 Sesto F.no (Firenze), Italy \\
$^{10}$Ioffe Institute for Physics and Technology, 194021 St.~Petersburg, Russia \\
$^{11}$Saint-Petersburg Polytechnic University, 195251, St.~Petersburg, Russia \\
$^{12}$International Space Science Institute, Hallerstrasse 6, 3012 Bern, Switzerland \\
$^{13}$NASA Marshall Space Flight Center, ZP12, 320 Sparkman Drive, Huntsville, AL 35805, USA \\
$^{14}$DESY, Platanenallee 6, D-15738 Zeuthen, Germany }

\begin{abstract}
We report on {\em Chandra X-ray Observatory (CXO)} observations of the pulsar wind nebula (PWN) associated with PSR B0355+54 (eight observations with a 395 ks total exposure, performed over an 8 month period).  
We investigated the spatial and spectral properties of the emission coincident with the pulsar, compact nebula (CN), and extended tail. 
We find that the CN morphology can be interpreted in a way that suggests a small angle between the pulsar spin axis and our line-of-sight, as inferred from the radio data.  
On larger scales, emission from the $7'$  ($\approx2$ pc)  tail is clearly seen. 
We also found hints of two faint extensions nearly orthogonal to the direction of the pulsar's proper motion.  
The spectrum extracted at the pulsar position can be described with an absorbed power-law + blackbody model.
The nonthermal component can be attributed to magnetospheric emission, while the thermal component can be attributed to emission from either a hot spot (e.g., a polar cap) or the entire neutron star surface.
Surprisingly, the spectrum of the tail shows only a slight hint of cooling with increasing distance from the pulsar. 
This implies either a low magnetic field with fast flow speed, or particle re-acceleration within the tail.  
We estimate physical properties of the PWN and compare the morphologies of the CN and the extended tail with those of other bow shock PWNe observed with long {\sl CXO} exposures.
\end{abstract}


\section{INTRODUCTION}
Pulsar wind nebulae (PWNe) are sources of nonthermal X-ray emission and prominent sites of particle acceleration.  
In addition to parameters characterizing pulsar rotation and magnetic field, the PWN appearance (and possibly other properties, such as spectrum and radiative efficiency) depend on the pulsar velocity (Gaensler \& Slane 2006; Kargaltsev \& Pavlov 2008).     
If a pulsar moves through the interstellar medium (ISM) with a supersonic speed, the ram pressure exerted by the ISM balances the pulsar wind pressure, resulting in a bow-shaped shock with most of the wind particles being confined in the direction opposite to that of the pulsar's motion (thus forming a pulsar tail).  
X-ray and radio observations of fast-moving pulsars have shown that they are indeed often accompanied by extended tails and bright compact nebulae (CNe) in the vicinity of the pulsar, although sometimes only one of those two structures is seen and CN morphologies can vary dramatically.
X-ray tail lengths can significantly exceed the sizes of the CNe, extending for a few parsecs behind the pulsar (see Kargaltsev et al.~2008).  
Occasionally, faint and highly elongated X-ray structures oriented at large angles with respect to the pulsar proper motion direction have been seen in deep X-ray images of supersonic pulsars. 
Studying PWNe of this type can provide information about the energetics of the pulsar wind, flow properties (such as velocity and magnetization), particle acceleration sites, ISM properties and entrainment in the pulsar wind, the angle between the spin axis and the pulsar velocity, and the kicks that pulsars receive during the supernova explosion.

PSR B0355+54 (B0355 hereafter), located at a parallax distance of $d=1.04^{+0.21}_{-0.16}$ kpc (Chatterjee et al.~2004), is a middle-aged radio pulsar (characteristic age $\tau\approx 0.56$ Myr) with a spin-down energy loss rate $\dot{E}=4.5\times 10^{34}$ erg s$^{-1}$, and a period $P= 156$ ms (more parameters listed in Table \ref{tbl-parameters}). 
Originally discovered in radio, pulsar B0355+54 has been observed in X-rays with both the {\sl Chandra X-ray Observatory} ({\sl CXO}, ACIS-S detector) and {\sl XMM-Newton} in 2004 and 2002, respectively.
These observations have revealed the presence of a $\sim 30''$ (0.15 pc) compact nebula (CN) and a fainter tail of emission visible up to  $\sim 6'$ southwest of the pulsar.  
The tail extends in the direction opposite to that of the pulsar's proper motion, $\mu_\alpha \cos\delta = 9.20 \pm 0.18$ and $\mu_\delta = 8.17 \pm 0.39$ mas yr$^{-1}$ (measured with the VLBA by Chatterjee et al.\ 2004), which corresponds to a transverse velocity of $v_\perp = 61^{+12}_{-9}$ km s$^{-1}$.
For the pulsar, CN, and tail, the X-ray spectra have been found to fit an absorbed power-law (PL) model with photon indices $\Gamma_{\rm PSR}^{\rm CXO}=1.9\pm0.4$ and $\Gamma_{\rm CN}^{\rm CXO}=1.5\pm0.3$ by McGowan et al.~(2006), and $\Gamma_{\rm Tail}^{\rm XMM}=1.84\pm0.43$ by Tepedelenlio{\v g}lu \& {\"O}gelman (2007).

In this paper we present the results of much deeper \chan observations.  
In Section 2 we describe the observations and data reduction.  
In Section 3 we present the images and spectral fit results.  
In Section 4 we discuss and interpret our findings, and in Section 5 we present our conclusions.

\begin{deluxetable}{lc}
\tablecolumns{9}
\tablecaption{Observed and Derived Pulsar Parameters \label{tbl-parameters}}
\tablewidth{0pt}
\tablehead{\colhead{Parameter} & \colhead{Value} }
\startdata
R.A. (J2000.0), $\alpha$ & 03 58 53.71650  \\
Decl. (J2000.0), $\delta$ & +54 13 13.7273  \\
Epoch of position (MJD) & 52334  \\
Galactic longitude (deg) & 148.1900  \\ 
Galactic latitude (deg) & +0.8109  \\ 
Spin period, $P$ (ms) & 156  \\
Dispersion measure, DM (pc cm$^{-3}$) & 57.17  \\ 
Distance, $d$ (kpc) & $1.04_{-0.16}^{+0.21}$  \\ 
Velocity, $v_\perp$, (km s$^{-1}$) & $61_{-9}^{+12}$ \\ 
Distance from the Galactic plane, $z$ (kpc) & 0.015  \\ 
Spin-down power, $\dot{E}$ (10$^{34}$ erg s$^{-1}$) & 4.5  \\ 
Spin-down age, $\tau_{\rm sd} = P/(2\dot{P})$ (Myr) & 0.56  
\enddata
\tablenotetext{}{Parameters from Chatterjee et al.~(2004).}
\end{deluxetable}

\section{OBSERVATIONS \& DATA REDUCTION}
B0355+54 was monitored over 8 months (from 2012 November 11 to 2013 July 11) for a total of 395 ks (see Table \ref{tbl-obs} for details).  
The data were taken with the ACIS-I instrument on board \chan in the Very Faint timed exposure mode (3.2 s time resolution).  
We processed the data using the \chan Interactive Analysis of Observations (CIAO) software (version 4.6) and {\sl CXO} Calibration Data Base (CALDB) version 4.5.9.  
The photon energies were restricted to the 0.5--8 keV range for all images and the subsequent spectral analysis.  
CIAO's Mexican-hat wavelet source detection routine {\tt wavdetect} (Freeman et al. 2002) was used to detect the  X-ray sources in the field (to be excluded from the  spectral analysis of the PWN emission).  
All spectra were extracted using CIAO {\tt specextract} and fitted with XSPEC (ver. 12.8.0).  
In all spectral fits we used the XSPEC {\tt wabs} model with absorption cross sections from Morrison \& McCammon (1983). 
We also created exposure maps from each observation and produced a merged, exposure-map-corrected image using the {\tt merge\_obs} tool from CIAO.

To study possible variability of the CN and produce an accurate combined image, we corrected the astrometry of the individual observations using the CIAO routine {\tt reproject$\_$aspect}\footnote{http://cxc.harvard.edu/ciao/ahelp/reproject\_aspect.html}, which runs {\tt wcs$\_$match} and {\tt wcs$\_$update} to improve aspect solutions using the list of point sources detected by {\tt wavdetect} as reference sources.  
The residual error of the image alignment is $<0\farcs2$, based on the comparison of the pulsar position in the aligned images. 
At a transverse speed of $v=61$ km s$^{-1}$, over the course of our observations (see Table 2), the pulsar would have traveled less than $0\farcs01$, which would not be detectable.
Relativistic flow in the PWN itself, however, might be seen with shifts of $1\farcs5-2\farcs6$ corresponding to velocities of $0.3c-0.5c$ at $d=1.04$ kpc\footnote{Projected velocities of up to $0.5c$ have been measured in the Crab PWN (Weisskopf et al.\ 2000), and evidence for blobs moving with $v_{\perp}\approx0.3c$ has been recently reported for PSR J1509--5850 (Klingler et al.\ 2016).}. 

To model and study the morphology of the emission in the immediate vicinity of the pulsar we used the {\sl Chandra} Ray Trace (ChaRT\footnote{http://cxc.harvard.edu/chart/}) and MARX software packages.
We simulated the \chan point spread functions (PSF) using the measured pulsar spectrum and ChaRT for each observation.  
The ChaRT output was supplied to MARX\footnote{http://space.mit.edu/CXC/MARX/} (ver.~5.1) to produce simulated event files, which were then merged.  
We used an Aspect Blur value of $0\farcs19$, as recommended by the \chan  X-ray Center\footnote{http://cxc.harvard.edu/chart/threads/marx/}.

\begin{figure}
\epsscale{1.1}
\plotone{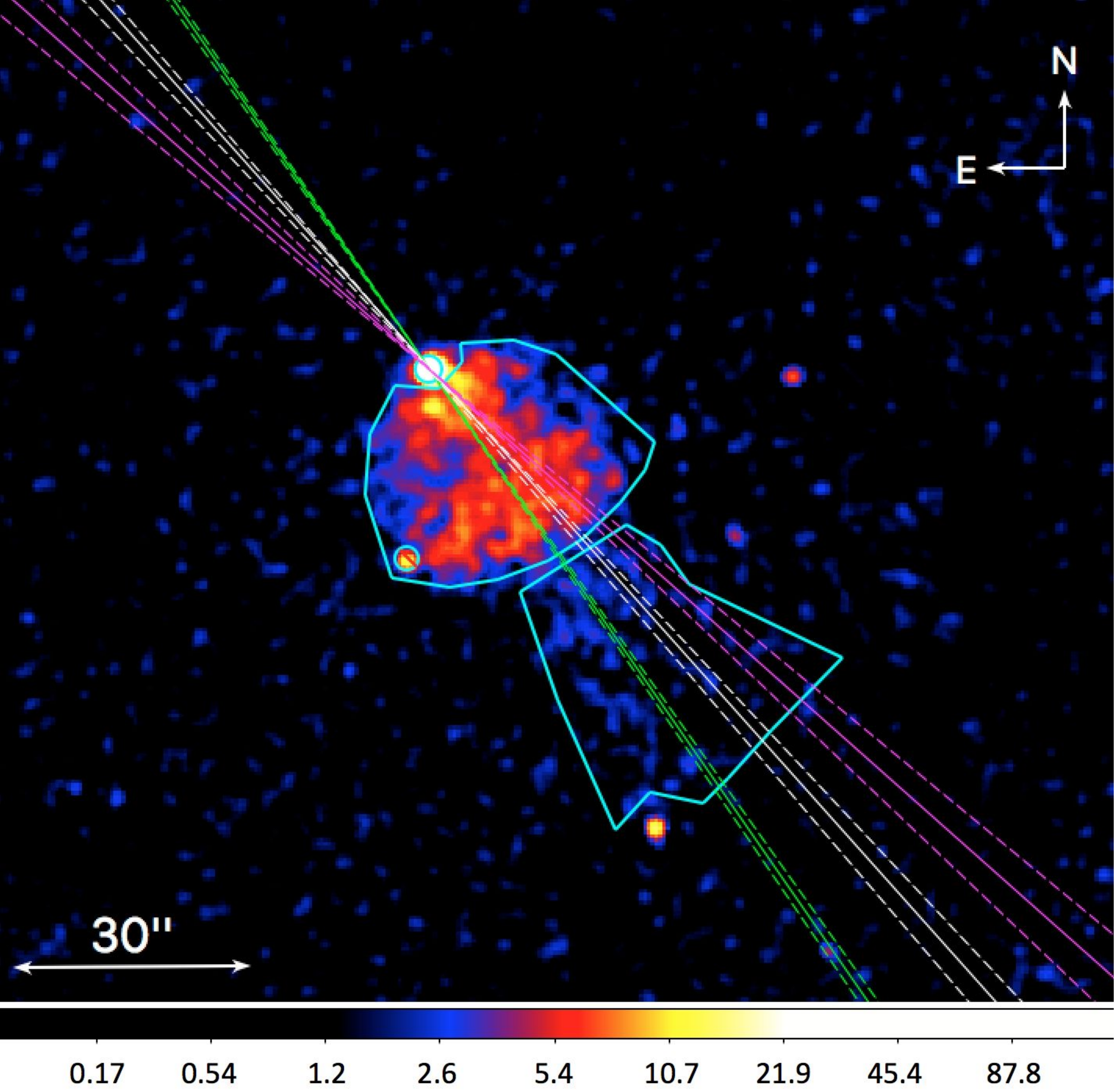}
\caption{Merged exposure-corrected ACIS image (0.5--8.0 keV, smoothed with an $r=1\farcs5$ Gaussian kernel) with the following spectral extraction regions shown: pulsar/core ($r=1\farcs7$ circle), compact nebula (CN) region (northern cyan polygon), and the mushroom ``stem'' region (southern cyan polygon).  The magenta lines represent the direction of proper motion ($48.4^{\circ} \pm 2.1^{\circ}$ East of North) as measured by Chatterjee et al.\ (2004), the white lines represent the direction of proper motion after the effects of solar motion and Galactic rotation have been removed ($41.8^{\circ} \pm 1.3^{\circ}$; Hobbs et al.\ 2005), and the green lines represent the measured symmetry axis ($34.7^{\circ} \pm 0.6^{\circ}$) of the CN (see below).  The point source that appears near the southeastern trailing edge of the CN (shown by the crossed circular region) was excluded from the CN spectral analysis.  The colorbar is in units of counts arcsec$^{-2}$}.
\label{image-regions}
\end{figure}

\begin{figure}
\epsscale{1.1}
\plotone{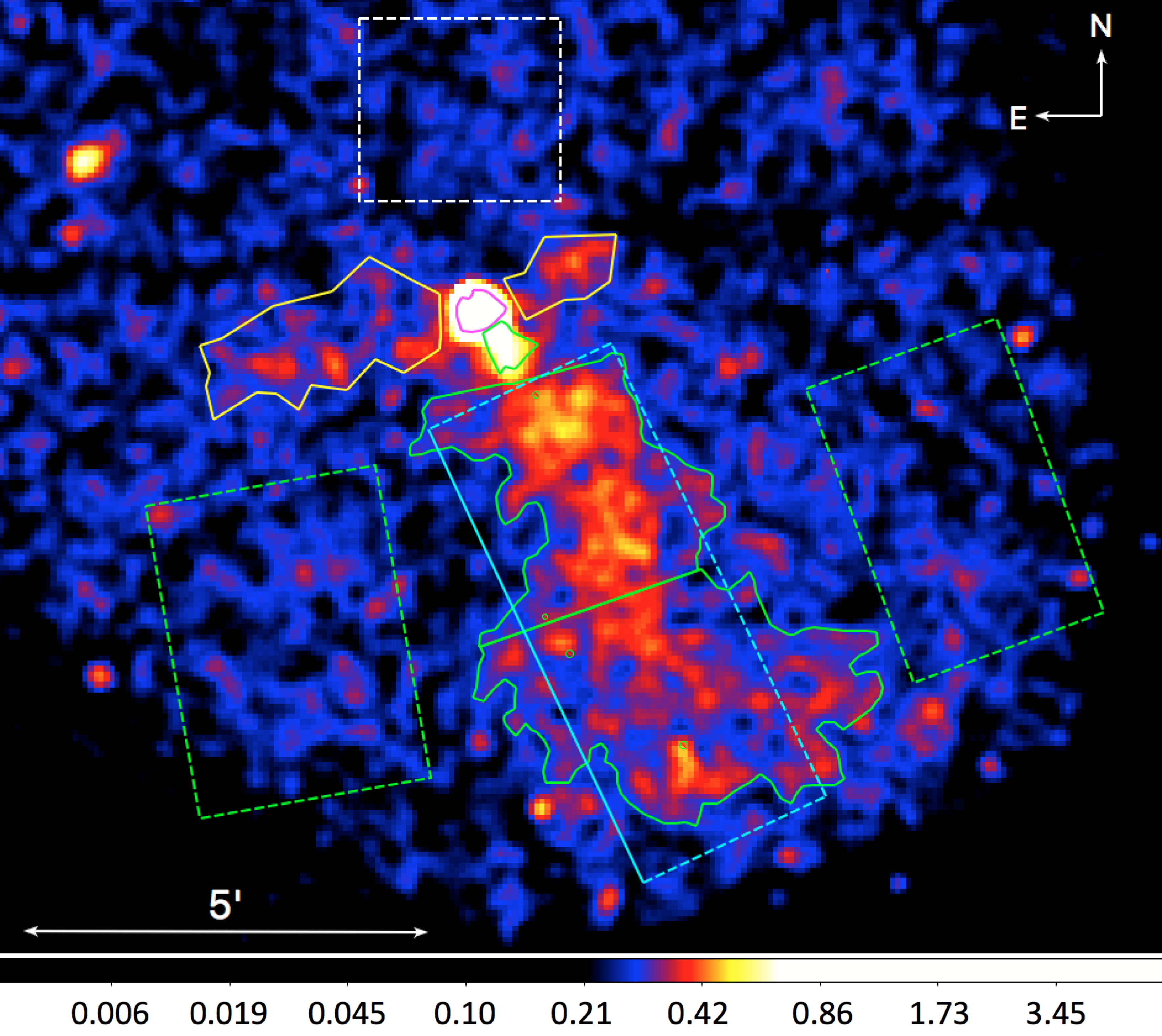}
\caption{Merged 0.5--8 keV exposure-corrected image produced from all 8 ACIS-I observations with point sources (detected by {\tt wavdetect}) removed.  The image is binned by a factor of 8 ($3\farcs9$ pixel size), and smoothed with an $r=11\farcs8$ Gaussian kernel.  The following spectral extraction regions are shown: the tail (green contour region; the line separates the ``near'' and ``far'' halves), and the ``whiskers'' (yellow polygons on either side of the CN).  The two green dashed rectangles to the sides of the tail are the background regions used in the tail's spectral fits.  The cyan $2\farcm5 \times 6\farcm2$ rectangular region (along the tail contour) is used to extract the tail brightness profile (see Figure \ref{tail-brightness-profile}).  The CN (magenta polygon) and stem (green polygon) are also shown (same regions as in Figure 2), along with the background region (dashed white box) used in the pulsar, CN, stem, and whiskers spectral fits.  The level for the green contour was chosen in such a away that it conforms to the tail morphology at the contrast shown.  The color bar is in units of counts arcsec$^{-2}$.}
\label{image-regions-tail-2}
\end{figure}

\section{RESULTS}

\subsection{PWN Spatial Morphology }
In our analysis we divided the PWN into five regions shown in Figures \ref{image-regions} and \ref{image-regions-tail-2}: (1) the pulsar (or ``core'') region within an $r=1\farcs7$ circle centered on the brightest pixel of the CN, (2) the CN within the region that resembles a ``mushroom cap'' (excluding the core region),  (3) the ``mushroom stem'' region which encompasses the faint extensions protruding from the CN, (4) the long, extended tail region (which we divided in two to search for spectral changes), and (5) the two ``whiskers'' composed of faint emission roughly orthogonal to the extended tail.
The CN region has a well-defined boundary due to the sharp brightness contrast (see Figure \ref{image-regions}). 
The extended tail is clearly seen only in the merged heavily binned (and smoothed) image (Figure \ref{image-regions-tail-2}).  
The boundary of the extended tail region was defined based on the surface brightness contour shown in Figure \ref{image-regions-tail-2}.
All field point sources detected by {\tt wavdetect} were excluded from the extended tail region.  
We also defined several background regions used to subtract the background in spectral fits (also shown in Figure \ref{image-regions-tail-2}).

\subsubsection{Pulsar Vicinity}

\begin{figure*}
\epsscale{1.1}
\plotone{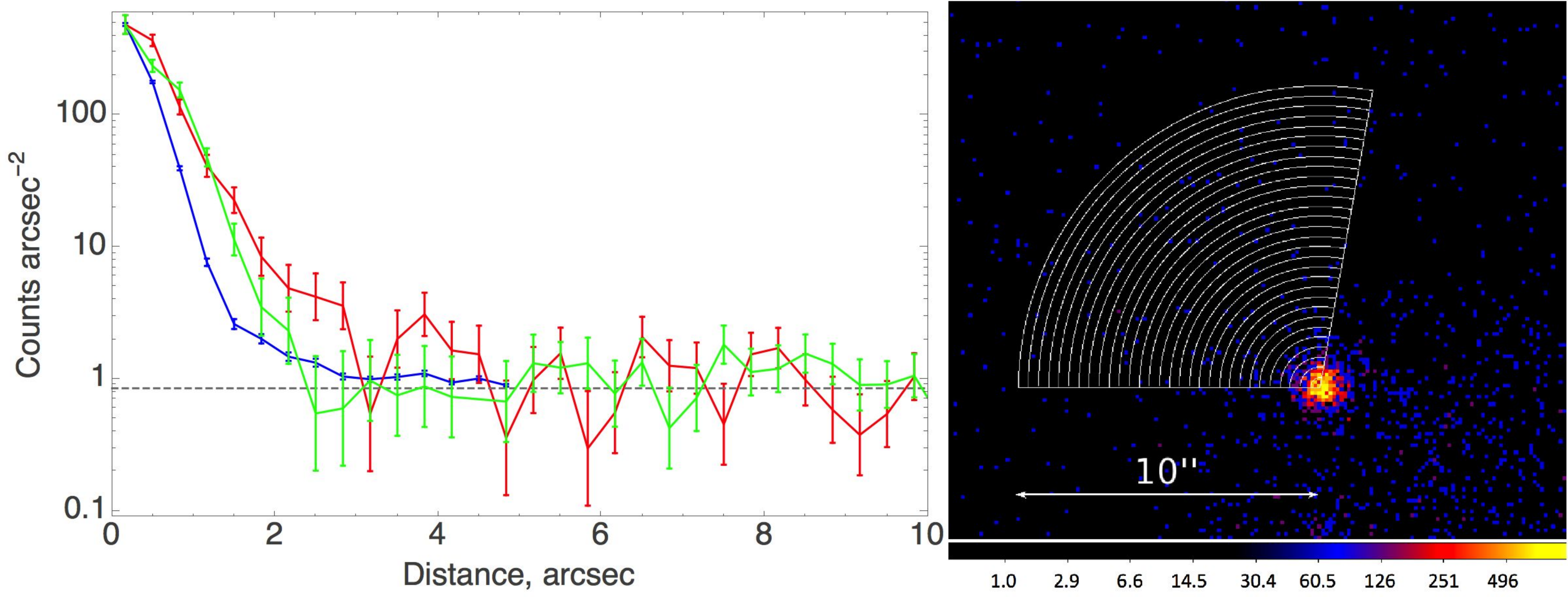}
\caption{Comparison of the MARX simulated point source radial profile (blue error bars), the nearby field point source radial profile (CXOU J035905.2+541455; green error bars), and the observed count distribution (red error bars) in the area ahead of the moving pulsar (shown in the right panel).  All profiles are normalized to the same value at the origin.  The average background surface brightness ($0.854\pm0.014$ counts arcsec$^{-2}$) shown by the dashed horizontal gray line was added to the simulated point source plot (blue).   The simulated point source (blue curve) appears to under-predict the actual width of the PSF (compared the green curve), thus we do not consider the MARX simulations to be reliable for the purpose of our analyses.  
The color bar (right panel) is in units of counts arcsec$^{-2}$.}
\label{wedge}
\end{figure*}

We begin our analysis by studying the pulsar's vicinity.  
This requires modeling and subtracting a point (unresolved) source at the location of the pulsar.  
We simulated point source images (produced for each observation because of the differing roll angles and slightly different off-axis angles; see Table \ref{tbl-obs}), combined them, and then subtracted them from the merged image of real data.   
However, the simulated PSF (the blue curve in Figure \ref{wedge}) under-predicts the width of the PSF for both B0355 and CXOU J035905.2+541455, a nearby point source used for comparison.  
Therefore, we do not consider the MARX simulation to be reliable for the purpose of our analysis here.

\subsubsection{Compact Nebula (CN)}
The deep merged images produced from all eight ACIS-I observations (Figures \ref{image-regions} and \ref{image-regions-tail-2}) reveal the CN  morphology, which can be described as a filled dome (or a mushroom cap when viewed together with the faint extension -- the mushroom ``stem'').  
The central region of the CN appears brighter than the sides.  
There is a clearly defined trailing edge beyond which the CN brightness falls off very abruptly.  
The sharp trailing edge is not straight and has a noticeable curvature to it.  

To measure the direction of the symmetry axis of the PWN, we calculated the sum of CN counts on both sides of the green line (shown in Figure \ref{image-regions}), which is rotated around the pulsar position until the number of counts is the same for each side.  
The 1$\sigma$ uncertainties in the direction of the symmetry axis (dashed green lines) correspond to the 1$\sigma$ deviation between the numbers of counts in the two parts of CN.

\begin{figure}
\epsscale{1.1}
\plotone{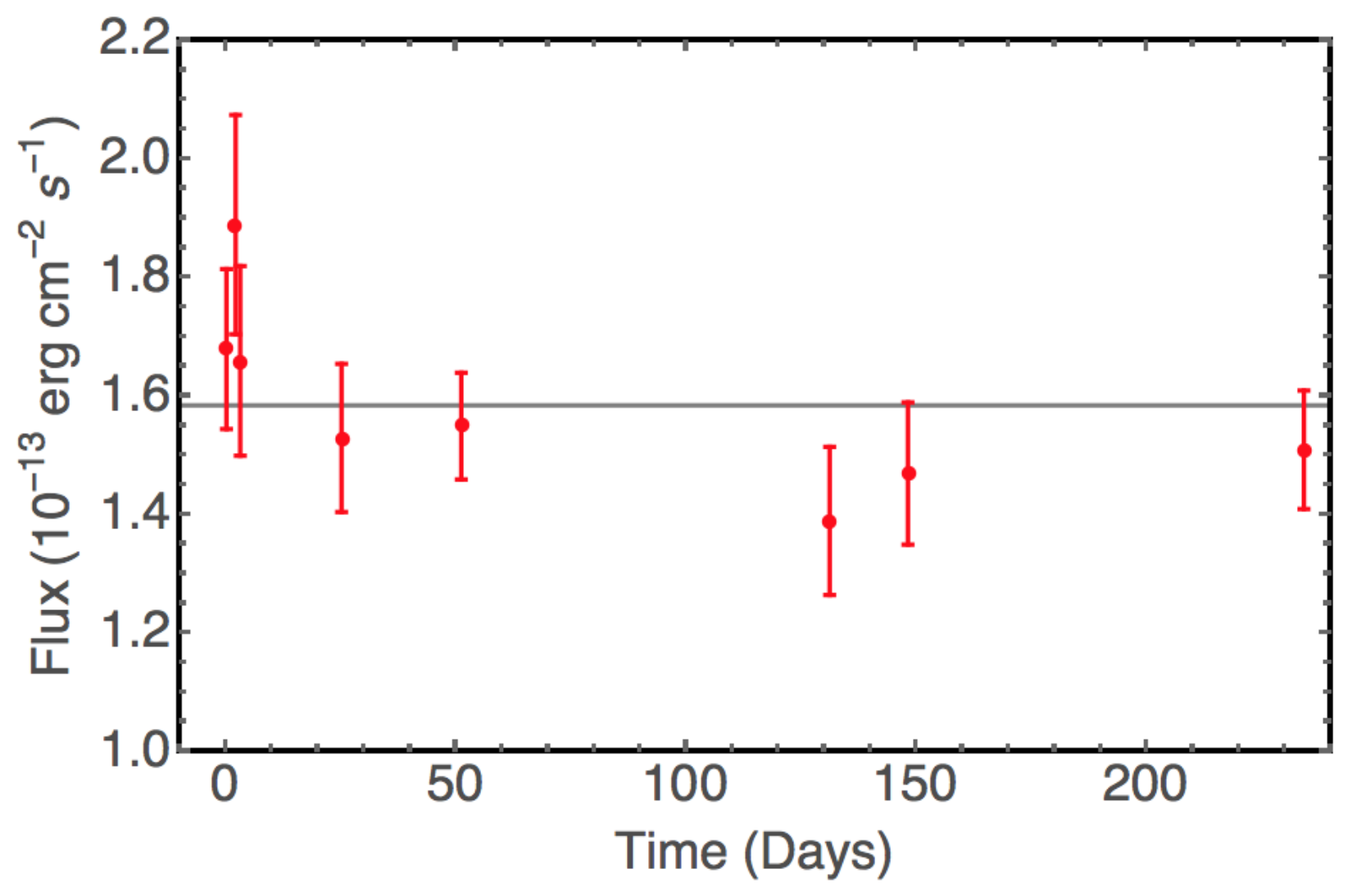}
\caption{Unabsorbed fluxes (0.5--8 keV, see Table \ref{tbl-obs}) of the individual fits of the CN region (shown in Figure \ref{image-regions}).  The gray line represents the average of the fluxes, $(1.59\pm0.03)\times10^{-13}$ erg cm$^{-2}$ s$^{-1}$.  These results are comparable to the CN flux from the archival ACIS-S observation (ObsID 4657), $(1.40\pm0.07)\times10^{-13}$ erg cm$^{-2}$.}
\label{image-mushroom-flux}
\end{figure}

We compared the individual images of the CN to look for possible variability.  
The flux for the entire CN as a function of time is plotted in Figure \ref{image-mushroom-flux}.  
Fitting the resulting lightcurve with a constant, $(1.59\pm0.03)\times10^{-13}$ erg cm$^{-2}$ s$^{-1}$, (the average of the individually-fitted fluxes listed in Table \ref{tbl-obs}) yields a reduced $\chi^2 =1.1$ for $\nu=7$ degrees of freedom (dof)), consistent with the lack of significant variability.
We also examined the structural changes in the subsequent CN images and could not identify any systematic variations in the CN structure.

\begin{deluxetable*}{cccccccc}[H]
\tablecolumns{9}
\tablecaption{Compact Nebula Data
\label{tbl-obs}}
\tablewidth{0pt}
\tablehead{
\colhead{ObsId} & \colhead{Date} & \colhead{Exposure} & \colhead{$\theta$} & \colhead{Net  Counts} & \colhead{Net  Count Rate} & \colhead{$\Gamma$} & \colhead{$F_{X,-13}  $}  \\
\colhead{} & \colhead{} & \colhead{ks} & \colhead{arcmin} &  \colhead{} & \colhead{ $10^{-3}$~s$^{-1}$} & \colhead{} & \colhead{ }  }
\startdata
14688 & 2012-11-19 & 26.62 & 1.02  & $196\pm14$ & $7.40\pm 0.54$ & $1.31_{-0.18}^{+0.19}$ & $1.68_{-0.13}^{+0.14}$ \\
15585 & 2012-11-21 & 21.78 & 0.98  & $166\pm13$ & $7.64\pm0.61$ & $1.48_{-0.18}^{+0.18}$ & $1.89_{-0.18}^{+0.19}$ \\
15586 & 2012-11-22 & 20.21 & 0.95 & $150\pm12$ & $7.42\pm0.63$ & $1.36_{-0.21}^{+0.21}$ & $1.66_{-0.15}^{+0.17}$ \\
14689 & 2012-12-14 & 67.19 &  2.33  & $460\pm21$ & $6.84\pm0.33$ & $1.66_{-0.14}^{+0.14}$ & $1.53_{-0.11}^{+0.14}$ \\
14690 & 2013-01-09 & 62.24 & 2.72 & $432\pm21$ & $6.94\pm0.35$ & $1.48_{-0.14}^{+0.14}$ & $1.55_{-0.06}^{+0.12}$ \\
15548 & 2013-03-30 & 65.90 & 1.73 & $338\pm18$ & $5.13\pm0.29$ & $1.65_{-0.14}^{+0.15}$ & $1.39_{-0.11}^{+0.14}$ \\
15549 & 2013-04-16 & 68.18 & 0.68 & $402\pm20$ & $5.89\pm0.31$ & $1.60_{-0.15}^{+0.15}$ & $1.47_{-0.11}^{+0.13}$ \\
15550 & 2013-07-11 & 63.00 & 1.59 & $431\pm21$ & $6.89\pm0.38$ & $1.40_{-0.14}^{+0.14}$ & $1.51_{-0.09}^{+0.11}$
\enddata
\tablecomments{$\theta$ is the angular distance between the pulsar and the optical axis of the telescope.  All errors listed in this table and paper are 1$\sigma$ errors, unless otherwise noted.  For spectral fits, a minimum of 16 counts per bin was required.  Counts and count rates are restricted to the 0.5--8.0 keV energy range.  For the above fits, $\Gamma$ was left free and untied, while $N_{\rm H}$ was left free and tied for all 8 observations. The best-fit $N_{\rm H}=(0.6\pm 0.1)\times10^{22}$ cm$^{-2}$ and the reduced $\chi^2 = 0.93$ (for 144 d.o.f.).  $F_{X,-13}$ is the unabsorbed flux in the 0.5--8 keV range in units of $10^{-13}$ ergs s$^{-1}$ cm$^{-2}$.}
\end{deluxetable*}

\vspace{0.4cm}
\subsubsection{Mushroom Stem}
A fainter emission protruding from the trailing edge of the CN in the direction opposite to that of the pulsar's proper motion (the ``stem'' of the mushroom) is visible up to $\sim30''$.  
The stem is clearly resolved in the transverse direction, becoming wider (but fainter) with the increasing distance from the pulsar.
The stem region is (on average) about a factor of 6 dimmer (in terms of surface brightness) than the CN. 
It may be composed of two structures that originate at the pulsar and slightly diverge from each other (with a half-angle of about $7^\circ$) further away, accounting for the apparent broadening of the stem with distance from the pulsar. 
If the stem indeed originates from the pulsar, it could account for the enhanced surface brightness along the symmetry axis of the CN.   
The stem is also seen in the first short \chan observation, although it is not as well resolved.

\subsubsection{Tail}
\begin{figure}
\epsscale{1.2}
\plotone{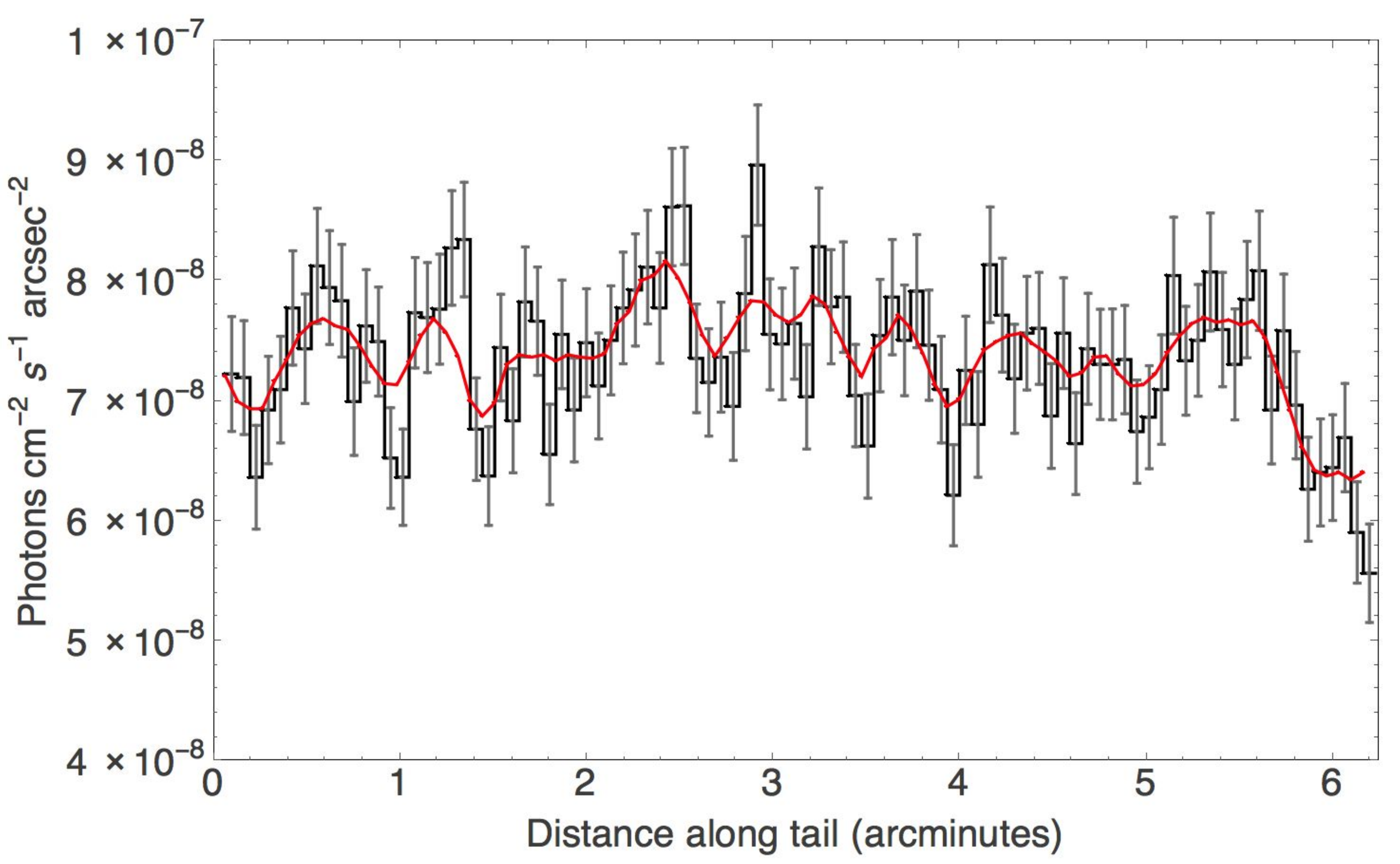}
\caption{The tail brightness profile as a function of distance from the pulsar, for the rectangular tail region shown in Figure 2.  The profile was obtained from the 0.5--8 keV exposure-map-corrected ACIS-I image binned by a factor of 8 (corresponding to a pixel size of $3\farcs9$).  The red curve represents the brightness profile when smoothed with a $11\farcs8$ Gaussian kernel).}
\label{tail-brightness-profile}
\end{figure}

A diffuse tail is discernible up to at least $7'$ southwest of the pulsar (limited by the ACIS-I field of view), in the direction roughly opposite to that of the pulsar's proper motion. 
The tail's average surface brightness (within the contour region shown in Figure \ref{image-regions-tail-2}) is dimmer than that of the CN by a factor of 40.  
The tail widens with distance from the pulsar and, for the part farthest from the pulsar, has a slightly lower average surface brightness (see Table 3).  
We show the tail's surface brightness as a function of distance from the pulsar in Figure \ref{tail-brightness-profile}.

\subsubsection{Whiskers}
Two very faint structures (``whiskers'') are discernable on either side of the CN, extending almost orthogonally to the direction of the tail and the pulsar's proper motion.  
Despite their faint appearance, 433 net (background-subtracted) counts were detected from the combined whiskers region, yielding a S/N ratio of 3.2.
Compared to the CN, the whiskers are dimmer by a factor of about 50, making them slightly fainter than the tail in terms of the average surface brightness.  
The whiskers do not appear to be bent back by the ram pressure caused by the pulsar's motion.
The eastern whisker appears to be approximately twice the length of the western one ($3\farcm5$ vs.\ $1\farcm8$, respectively).

\subsection{Spatially-Resolved Spectra}
\begin{figure}[h]
\epsscale{1.1}
\plotone{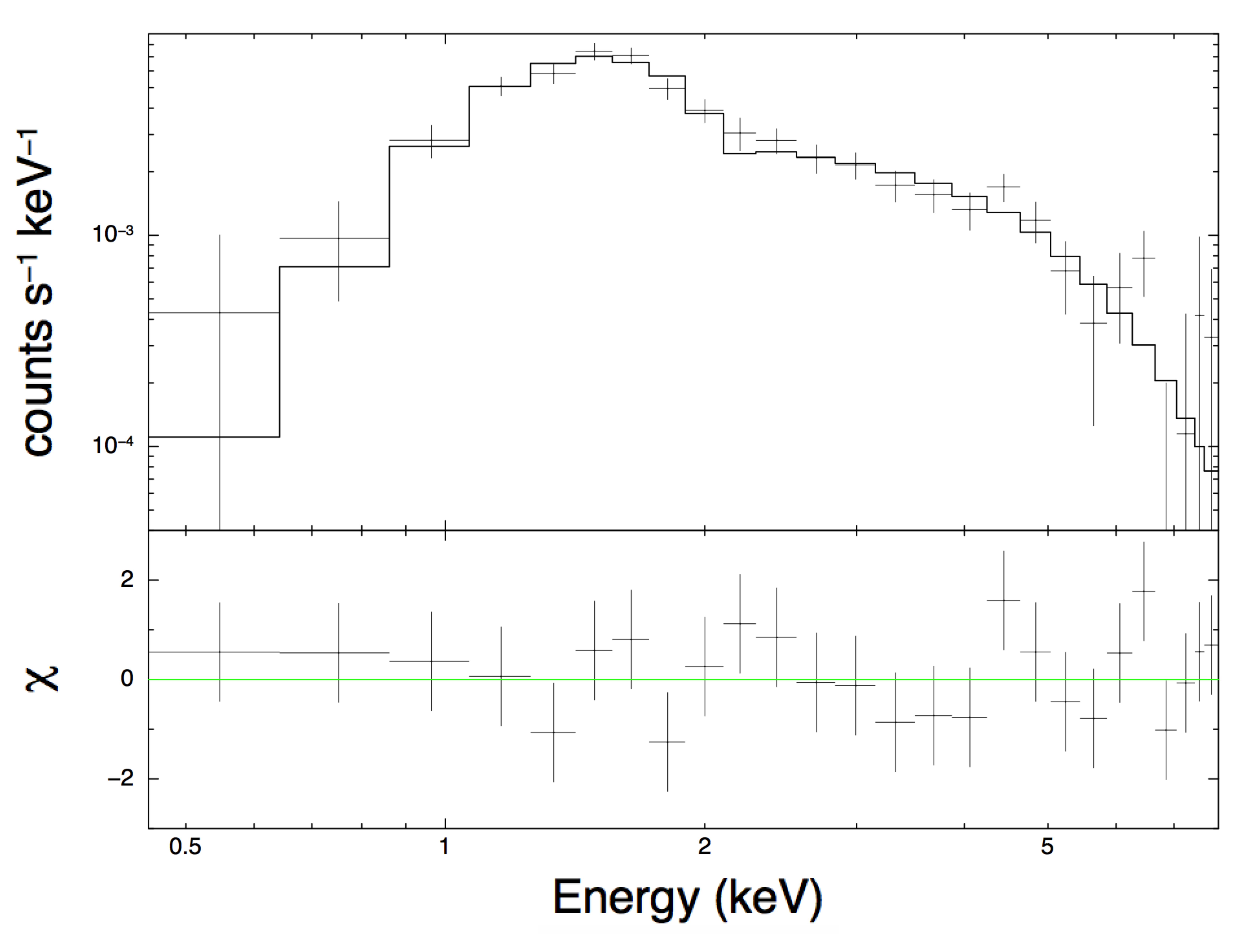}
\caption{The ACIS-I spectrum for the entire tail obtained by merging the spectra from 8 observations. The best fit shown corresponds to an absorbed PL with $\Gamma=1.74\pm0.08$ and  $N_{\rm H,22}=0.61$. The merged spectrum is shown for illustrative purposes only.   See Table \ref{spatially-resolved-spectra} for details.}
\label{tail-spectra}
\end{figure}

We extracted the spectra of the core/pulsar, CN, stem, whiskers, and the extended tail for the eight observations listed in Table \ref{tbl-obs} and fitted them.  
We also split the tail into two sub-regions: the near-half and far-half (as shown in Figure \ref{image-regions-tail-2}).  
All spectra were fitted satisfactorily with an absorbed power-law (PL) model, except for the pulsar/core region, which is best fit with a PL+blackbody (PL+BB) model.  
Fitting the CN spectrum yields the Hydrogen column density $N_{\rm H} = (0.61\pm0.09) \times 10^{22}$ cm$^{-2}$ (a comparable, albeit less certain value is obtained from the fit of the tail spectrum)\footnote{Alternatively, fitting together the pulsar, CN, and tail regions simultaneously with a tied $N_{\rm H}$ and free $\Gamma$ values and normalizations gives $N_{\rm H}=(7.5\pm0.6)\times10^{21}$ cm$^{-2}$, which is within 1.6$\sigma$ of the $N_{\rm H}$ value for the CN (see Table \ref{spatially-resolved-spectra}.  The results from this approach do not significantly differ from those listed in Table \ref{spatially-resolved-spectra}.  Therefore, we use the $N_{\rm H}$ value obtained from the CN fit, as it had the best constrained $\Gamma$).}.
We fix the $N_{\rm H}$ at this value while fitting the spectra of other regions (see Table \ref{spatially-resolved-spectra}) because of the poorer statistics. 
For each of the regions, we simultaneously fit the data from all eight observations used (see Table \ref{spatially-resolved-spectra}).
The photon indices for the pulsar/core and the CN are $\Gamma_{\rm PSR}=1.85 \pm 0.12$ and $\Gamma_{\rm CN}=1.54\pm 0.05$, respectively.  
The indices are virtually the same for both halves of the tail: $\Gamma_{\rm T,n} = 1.72\pm 0.10$ for the near-half of the tail, and $\Gamma_{\rm T,f} = 1.77\pm 0.11$ for the far-half, while $\Gamma_{\rm T} = 1.74\pm 0.08$ for the entire tail.  
For the whiskers, $\Gamma_{\rm W} = 1.60\pm0.31$ with  $\chi_{\nu}^2=1.32$ for $\nu=$33 d.o.f., and for the stem, $\Gamma_{\rm Stem} = 1.73\pm0.17$ with  $\chi_{\nu}^2=0.82$ for $\nu=$59 d.o.f.
For illustration, the merged tail spectrum is shown in Figure \ref{tail-spectra}.

Our spectral measurements are more accurate than those obtained from the earlier \chan  and {\sl XMM-Newton} observations.  
As a result, we can better constrain possible spectral changes due to synchrotron cooling. 
We observe only a hint of softening, $\Delta\Gamma=0.2 \pm 0.1$, between the CN and pulsar tail, while virtually no spectral slope change is seen between the two halves of the tail.  
We discuss the implications of these findings in Section 4.

The softer spectrum and rather poor quality of the pulsar/core region PL fit ($\chi_\nu^2=1.37$) suggest a possible contribution of thermal emission from the neutron star (NS) surface, which can be approximated by a blackbody (BB) model.  
A PL+BB model provides  a better quality fit ($\chi_\nu^2=1.14$), yielding $\Gamma_{\rm PSR}\approx1.45$, $kT=0.16~(0.12, 0.21)$ keV, and the projected emitting area $\mathcal{A}= 0.20~(0.03, 3)$ km$^2$, where the first and second number in parentheses correspond to the $1\sigma$ lower and upper bounds of the parameter.  
The improvement in the fit quality due to the addition of the BB component is significant at the 97.6\% level according to the F-test. Figure \ref{contour} shows the contribution of the thermal component at lower energies (left panel) and constraints on the temperature and size of the emitting region (right panel).

\begin{deluxetable*}{lccccccccc}[H]
\tablecolumns{4}
\tablecaption{Spatially-Resolved Spectra of PWN Regions
\label{spatially-resolved-spectra}}
\tablewidth{0pt}
\tablehead{\colhead{Region} & \colhead{Area} & \colhead{ $I$ } & \colhead{$\Gamma$} & \colhead{$\mathcal{N}_{-5}$} & \colhead{ $\mathcal{B}_{-8}$ } & \colhead{Reduced $\chi^2$} & \colhead{$\log F_X$} & \colhead{$L_X$} & \colhead{$\log \eta_X$} \\ \colhead{} & \colhead{arcsec$^2$} & \colhead{cts arcsec$^{-2}$} & \colhead{} & \colhead{} & \colhead{} & \colhead{(dof)} & \colhead{} & \colhead{$10^{31}$ erg s$^{-1}$} & \colhead{} }
\startdata
Pulsar/core  & 9.44 &  $80.8\pm2.9$ & $1.85_{-0.12}^{+0.12}$ & $0.83\pm0.09$ & 88 & 1.37 (32) & $-13.38\pm 0.02$ & 0.54 & $-3.92$ \\
Pulsar/core\tablenotemark{a} & 9.44 & $80.8\pm2.9$ & $1.45_{-0.24}^{+0.21}$ & $0.51\pm0.16$ & 54 & 1.14 (30) & $-13.21\pm0.03$ & 0.80 & $-3.75$ \\
CN & 791 & $3.45\pm0.07$ & $1.54_{-0.05}^{+0.05}$ & $2.2\pm0.10$ & 2.8 & 0.96 (159) & $-12.85\pm 0.01$ & 1.83 & $-3.39$ \\
Whiskers & 14726 & $0.067\pm0.008$ & $1.60_{-0.29}^{+0.32}$ & $0.48\pm0.12$ & 0.03 &  1.32 (33) & $-13.44\pm 0.06$ & 0.47 & $-3.98$ \\
Stem & 824 & $0.55\pm0.03$ &  $1.73_{-0.16}^{+0.17}$ & $0.36\pm0.05$ & 0.44 & 0.82 (59) & $-13.70\pm 0.03$ & 0.26 & $-4.24$ \\
Tail (Entire) & 59041 & $0.087\pm0.004$ & $1.74_{-0.08}^{+0.08}$ & $5.4\pm0.36$ & 0.091 & 1.13 (135) & $-12.53\pm 0.02 $ & 3.8 & $-3.07$ \\
Tail (Near) & 24223 & $0.11\pm0.004$ & $1.72_{-0.10}^{+0.10}$ & $2.3\pm0.21$ & 0.095 & 1.38 (43) & $-12.90\pm 0.02$ & 1.6 & $-3.44$ \\
Tail (Far) & 34818 & $0.070\pm0.004$ & $1.77_{-0.11}^{+0.11}$  & $3.1\pm0.28$ & 0.089 &1.15 (69) & $-12.79\pm 0.02$ & 2.1 & $-3.33$ \\
\tablecomments{Spectral fits for the extraction regions shown in Figures 2 and 3 with an absorbed PL model.  The $N_{H}$  is fixed at $6.1\times 10^{21}$ cm$^{-2}$, which is the best-fit value for the CN and tail spectra (when fitted independently).  The net surface brightness $I$ (in cts arcsec$^{-2}$) is listed for each region.  $\mathcal{B}=\mathcal{N}/A=10^{-8}\mathcal{B}_{-8}$ is the average spectral surface brightness at $E=1$ keV (photons s$^{-1}$ cm$^{-2}$ keV$^{-1}$ arcsec$^{-2}$), where $\mathcal{N}=10^{-5}\mathcal{N}_{-5}$ is the normalization of the photon spectral flux from the PL fits (photons s$^{-1}$ cm$^{-2}$ keV$^{-1}$) measured in area $A$.  Unabsorbed flux $F_X$, luminosity $L_X$, and X-ray efficiency $\eta_X=L_X/\dot{E}$ are for the distance of $1.04$ kpc and correspond to the 0.5--8 keV range.}
\tablenotetext{a}{The  PL component of the PL+BB fit to the spectrum of the pulsar/core region.}
\enddata
\end{deluxetable*}

\section{DISCUSSION}
As described above, the PWN morphology is quite complex, with feature scales varying from arcseconds to arcminutes. 
Overall, the X-ray images are dominated by the PWN (i.e., all regions except the pulsar/core), whose luminosity in 0.5--8 keV, $L_{\rm PWN}=6.4\times10^{31}$ erg s$^{-1}$, exceeds that of the unresolved pulsar/core by a factor of $13$.  
Below we discuss and attempt to interpret the individual features of the PWN.

\subsection{Core/Pulsar}
\begin{figure*}
\epsscale{1.1}
\plotone{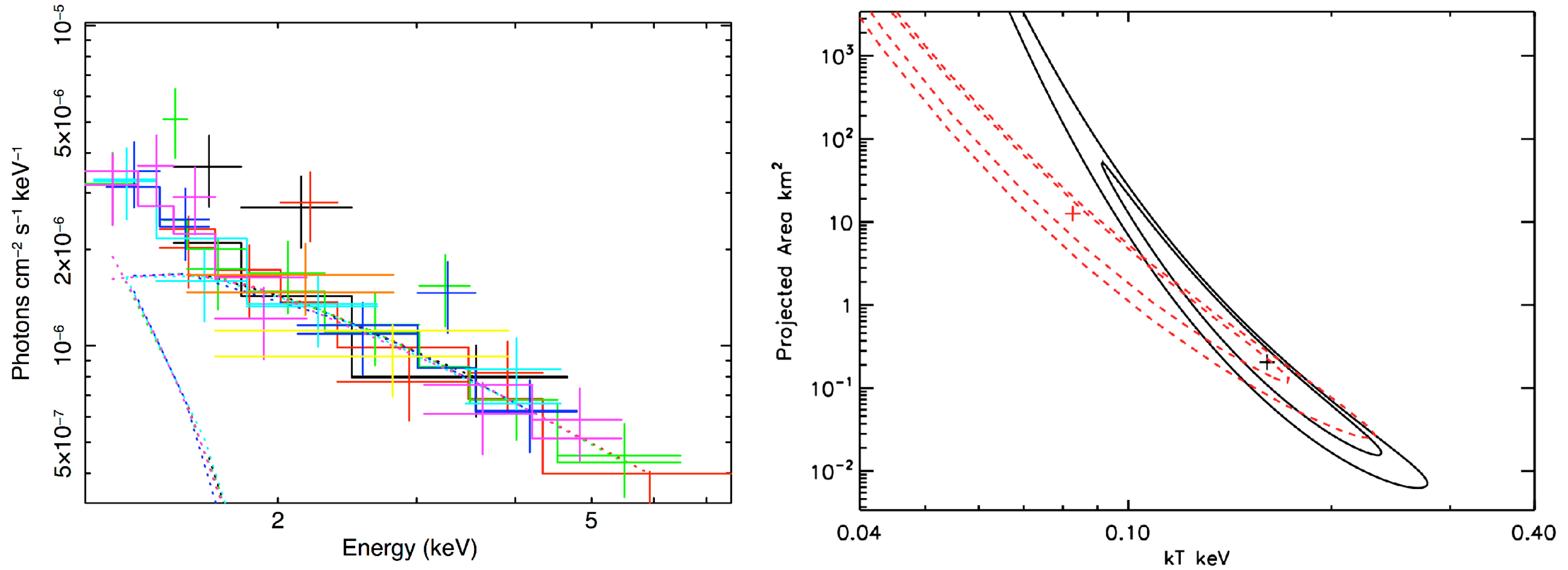}
\caption{{\sl Left:} Plot of the unfolded  spectrum fitted with a PL+BB model.   {\sl Right:}   Contour plot of the  projected emitting BB area  versus  temperature for the PL+BB fit (black solid line) and a similar plot for the PL+NSA fit to the pulsar/core region (red dashed line; 1$\sigma$ and  2$\sigma$ confidence contours are shown for both plots).}
\label{contour}
\end{figure*}

The superior quality of the PL+BB fit (compared to the PL fit) suggests the presence of a soft thermal component in addition to the PL whose slope is comparable with the CN slope. 
The large uncertainties of the thermal component and, particularly, the poorly constrained size of the emitting area (see Figure \ref{contour}) do not allow us to make a firm conclusion on whether the thermal component comes from the bulk of the NS surface or from a hot polar cap.  
If the emission comes from the surface of an $R_{\rm NS}=13$ km NS surface, the fit implies $T\sim0.8$ MK, which is hotter then what is expected from standard cooling at B0355's spin-down age \citep{2004ARA&A..42..169Y} and falls outside the $1\sigma$ contour shown in Figure \ref{contour}. 
The temperature would be lower, $\sim0.6$ MK, if the NS (with the same radius) has a hydrogen atmosphere (NSA model; \citealt{1995ASIC..450...71P}).  
Alternatively, the thermal component could be attributed to a hot ($T\sim 1.8$ MK) spot with the size ($R=\sqrt{\mathcal{A}/\pi}\sim 250$ m), comparable to the size of a pulsar's polar cap, $R_{\rm PC}=(2\pi R_{\rm NS}^3 / cP)^{1/2}\approx 540$ m.

Although the emission from the pulsar/core appears to be centrally peaked, we cannot claim with absolute confidence that it represents the pulsar emission until pulsations are detected\footnote{A search for pulsations in our data is not possible because the ACIS-I data was taken in Full Frame mode, which has a coarse 3.2 s timing resolution.  The timing analysis of the {\sl XMM-Newton} data by McGowan et al.\ (2008) produced only marginal detection of a pulsed signal, likely because of substantial contamination by CN photons.}. 
It is unlikely that the termination shock is hidden inside the unresolved core because this would imply an improbably large pulsar velocity (see estimates below).

\subsection{Compact Nebula (CN)}
\subsubsection{Bow Shock and Pulsar Velocity}

The cometary shape of the CN indicates that the pulsar wind flow is deflected by the ram pressure of the ISM.
Comparing the brightness profile of the region ahead of the pulsar with the radial profile of the field point source CXOU J035905.2+541455, we find that the emission at the CN apex becomes distinguishable from the background at a distance of $\approx3\farcs5$ in front of the pulsar (see Figure \ref{wedge}), which corresponds to the projected stand-off distance $r_{0,\perp} \simeq 5.4 \times 10^{16} d_{1.04}$ cm.
Balancing the ram pressure of the ISM with the isotropic pulsar wind pressure\footnote{Note that isotropic pulsar wind is assumed for simplicity.}, $\mu m_{\rm H}n v_p^2 = \dot{E}/(4\pi c r_{0}^2)$, where $m_{\rm H}$ is the mass of the Hydrogen atom, $\mu=1.3$ is used to convert H density to total density of the ambient medium, and $v_p=v_{p, \perp}/\sin i$ is the pulsar velocity, allows us to estimate the ISM number density $n=0.5 (\sin i)^2$ cm$^{-3}$ (where $i$ is the angle between the pulsar velocity and the line of sight).  
The exact inclination angle $i$ is not known; however, it can not be too different from $90^{\circ}$ given that the stand-off distance $r_{0,\perp}$ is small compared to the size of the CN.  
Based on the pulsar's Galactic coordinates ($l=148\fdg19$, $b=0\fdg81$) and the $d=1.04$ kpc distance, it is reasonable to assume that the ISM is in the ``warm'' phase (see p.~525 of Cox 2000), with sound speeds of $c_{\rm ISM} \sim8-13$ km s$^{-1}$ and the Mach number, $\mathcal{M}=v_p/c_s$,  is  (5--8)/$\sin i$, depending on the value of $c_s$ and the inclination angle $i$.

For highly supersonic motion and small values of the magnetization parameter of the initially isotropic pre-shock pulsar wind, the termination shock (TS) bullet's cylindrical radius $r_{\rm cyl} \simeq r_0$, while the lengths of the TS bullet is expected to be $l_{\rm bullet}\simeq 6 r_0$ (Gaensler et al. 2004; Bucciantini et al.\ 2005, hereafter B+05). 
This provides a poor match to the observed shape of the CN, which can be described as a dome rather than a bullet. 
In addition to the expected anisotropy of the pulsar wind, it is likely that the motion of the pulsar in this case is only mildly supersonic, however, low-$\mathcal{M}$ simulations (van der Swaluw et al.\ 2003) fail to reproduce the dome-shaped TS as well. 
It is reasonable to expect that the CN could become wider (compared to isotropic wind scenario) if the outflow is dominated by the equatorial component perpendicular to the pulsar velocity vector.
In this case, the toroidal structure associated with the TS\footnote{Note that the sizes of the toroidal structures seen in a number of PWNe (KP08) can be substantially larger than the TS shock size, e.g., in the Crab PWN, the torus is a factor of 3 larger than the size of the inner ring (Weisskopf et al.\ 2000).} will be deformed and pushed back by the ram pressure of the oncoming medium.  
Evidence for such a transitional deformation between the toroidal structure and cometary shapes is seen in the Vela PWN (see Figure 2 in Pavlov et al.~2003), the IC443 PWN (Swartz et al.~2015) and, perhaps, in the CTB~80 PWN (Moon et al.~2004).  

Since the CN is highly symmetric, we measured the orientation of the axis of symmetry (projected onto the sky plane) as $\phi\approx 34.7 \pm 0.6^\circ$ East of North (which coincides with the approximate axis of symmetry of the tail; see Section 3.1.2).
This angle is noticeably different ($\Delta\phi\approx7.1^\circ$) from the velocity vector's angle of $(41.8\pm1.3)^{\circ}$ East of North as reported by Hobbs et al.\ (2005), who removed effects of solar motion and Galactic rotation from the values obtained by Chatterjee et al.\ (2004) based on VLBA observations (see Figure 1). 
The mismatch could be explained by motion of the ISM, since the axis of symmetry should trace the relative speed of the pulsar with respect to the ISM, and not its absolute direction. 
If the mismatch is indeed attributed to ISM motion, it would imply a projected bulk ISM flow velocity $v_{\rm ISM}\approx8\ {\rm km\ s}^{-1}$ (directed southeast, perpendicular to the pulsar's direction of proper motion (the white line in Figure \ref{image-regions}).

\subsubsection{CN Morphology}
The CN of B0355 features a ``filled'' morphology with the central region of the CN being noticeably brighter than the sides (see Figure 1).  
Such an appearance is in stark contrast with the ``hollow'' morphologies of the Geminga and PSR J1509--5850 PWNe (Pavlov et al.~2010 and Klingler et al.~2016, respectively), but similar to that of the (filled) Mouse PWN (Gaensler et al.~2004).  
If the spin axis of B0355 is directed close to the line of sight, the brighter central region of the CN could represent the jets (launched along the NS spin axis) that are being swept back by the ram pressure of the oncoming ISM.
In this interpretation, the jets appear to be superimposed on the sky close to each other, but they separate slightly beyond the trailing edge of the CN, contributing to the split appearance of the stem (see Figures 1 and 2).  
Note that this scenario explains the nearly symmetric disposition of the split stem about the axis of symmetry.  
If axisymmetric sweepback provides the only perturbation of twin polar jets, then we can use the stem opening half-angle to infer that the angle between the proper motion and the {\sl projected} jet axis is $\sim 7^\circ$. 
For such a small angle, the jets would appear nearly straight, as observed, but unless the jet axis is projected very close to the proper motion axis, the full angle to the line of sight is small: $< 14^\circ$ (1 $\sigma$) and $< 44^\circ$ (90\% confidence level).

In fact, there is a good reason to believe that the spin axis of B0355 is nearly aligned with our line of sight, since it is not detected in the $\gamma$-ray band. 
Indeed, Romani et al.\ (2011) showed that B0355 is one of the ``subluminous'' $\gamma$-ray pulsars (objects fainter than expected from their spin-down parameters), and argued from radio emission properties that this is an orientation effect.
As shown by Figure 3 of Romani et al.\ (2011), the radio pulse properties imply a spin axis within 40$^\circ$ of the line-of-sight, where indeed $\gamma$-ray pulses would not be visible. 
Geminga and PSR J1509--5850, in contrast, are $\gamma$-loud and hence must be viewed nearly orthogonally to the spin axis. 
We thus expect that their swept-back jets would lie near the plane of the sky and are widely separated and highly curved\footnote{This is assuming that the spin and velocity vectors are substantially misaligned.}, as observed (Klingler et al.\ 2016; Posselt et al.\ in prep).

Note that while there is a statistical trend for spin and proper motion vectors to be aligned (e.g. Ng \& Romani 2007), there are examples of highly misaligned objects (e.g. the IC433 PWN and PSR B1706$-$44, Swartz et al.\ 2015 and Romani et al.\ 2005, respectively). 
One conclusion of Ng \& Romani (2007) is that, for the kick models studied, misalignment should be most common for pulsars with low space velocities. 
B0355's low velocity supports this hypothesis.

Orientation alone can hardly explain the contrasting morphologies of CNe observed in several of the PWNe mentioned above. 
For example, the Mouse PWN (Gaensler et al.\ 2004) and PSR J1741--2054's PWN (Auchettl et al.\ 2015) do not show any strong jet component, and may be dominated by equatorial outflows. 
Two pulsars with jets likely aligned with our line of sight, PSRs B0656+14 and B1055--52 (B{\^i}rzan et al.\ 2016; Posselt et al.\ 2015, respectively),  have very faint PWNe. 
Other factors, such as the spin-magnetic axis angle, and the angle between the spin axis and velocity vector, must control the relative brightness of the polar and equatorial wind components.

\subsection{Stem and Tail}
The visible (projected) length of the pulsar tail spans $7'$ (2.1 pc at $d=1.04$ kpc) behind the pulsar.  
It is possible that the actual length is larger, but it would then be outside the ACIS-I field of view.  
Beyond $\sim3'$ from the pulsar, the tail broadens and drops noticeably in surface brightness (see Figure \ref{image-regions-tail-2}). 
The near half of the tail (the one that is closer to the pulsar) appears to be more or less symmetric with respect to the CN symmetry axis, while the second half seems to bend $\sim 30^\circ$ south, and expand.
Although the symmetry axis of the PWN and tail may not exactly align with the proper motion direction in the case of anisotropic pulsar wind (Vigelius et al.\ 2007), such distortions caused by misalignment are not expected to manifest themselves at large distances from the pulsar.
Thus, the observed bending could be caused by ISM pressure nonuniformity, instabilities in the tail flow, or ISM entrainment, which brings the flow close to being at rest with respect to the ambient medium.   
A similar bending is seen in the tail of PSR J1741--2054; however, its morphology is somewhat different, as it appears to be composed of expanding bubble structures (Auchettl et al.\ 2015).
Despite the sharp drop in the surface brightness behind the trailing edge of the CN, there appears to be only modest (if any) spectral changes between the CN, the stem, and the tail. 
The spectrum softens by $\Delta\Gamma\lesssim0.2$ between the CN and the tail (see Table \ref{spatially-resolved-spectra} and Figure \ref{photon_index_distance}), suggesting that the sharp drop in brightness\footnote{In the Vela PWN, the brightness change between the arcs (presumably associated with the termination shock) is only a factor of 1.5--2 over $3''$-$4''$  length scale comparable to $1''$ at the B0355's distance.} (a factor of $\gtrsim6$ over a $5''$ distance) is not associated with rapid cooling but should rather be attributed to the rapid flow expansion or/and drop in magnetic field strength.   
Rapid expansion is expected in models that take into account ISM entrainment into the wind behind supersonically moving pulsars (see Figure 7 in Morlino et al.\ 2015).

A similarly small spectral evolution is seen between the compact PWN and tail of PSR J1509--5850 (Klingler et al.\ 2016).  
On the other hand, much more pronounced spectral softening trends have been measured in the N157B ($\Delta\Gamma\approx1.3$, Chen et al.\ 2006), the Lighthouse ($\Delta\Gamma\approx0.7$, Pavan et al.\ 2016), and the Mouse (Gaensler et al.\ 2004; $\Delta\Gamma\approx1.5$, Klingler et al.~in prep) PWNe.

\subsubsection{Physical Properties}
The magnetic field strength can be estimated from measurements of synchrotron surface brightness (see e.g., Pavlov et al.\ 2003).   
For a given magnetization,  $k_m=w_B/w_e$, where $w_B=B^2/(8\pi)$ is the magnetic energy density and $w_e$ is the energy density of relativistic particles, the magnetic field strength can be expressed as
\begin{equation} \footnotesize
B=7.2\left( \frac{k_m}{a_p(3-2\Gamma)} \left[ E_{M,p}^{(3-2\Gamma)/2} - E_{m,p}^{(3-2\Gamma)/2}\right] \frac{\mathcal{B}_{-8}}{\bar{s}_{17}} \right)^{2/7} {\rm \mu G.}
\end{equation}
In this equation $\mathcal{B}=\mathcal{N}/A=10^{-8}\mathcal{B}_{-8}$ is the average spectral surface brightness at $E=1$ keV (photons s$^{-1}$ cm$^{-2}$ keV$^{-1}$ arcsec$^{-2}$), $\mathcal{N}$ is the normalization of the photon spectral flux (from the PL fits) measured in area $A$, $\bar{s}=10^{17} \bar{s}_{17}$ cm is the average length of the radiating region along the line of sight, $E_{m,p}=E_m/y_{m,p}$, $E_{M,p}=E_m/y_{M,p}$, $E_m$ and $E_M$ are the lower and upper energies of the photon PL spectrum (in keV), and $y_{m,p}$, $y_{M,p}$ and $a_p$ are the numerical coefficients\footnote{See Table 2 in Ginzburg \& Syrovatskii (1965).} whose values depend on the slope, $p=2\Gamma-1$, of the electron  spectral energy distribution (SED).

Table \ref{spatially-resolved-spectra} lists $\Gamma$ and $\mathcal{B}$ for the two sections of the tail and the CN, and $\bar{s}$ can be assumed to be equal to the tail's width.
The magnetic field strength also depends on the boundary energies of the electron SED, which are poorly constrained because the tail is not detected outside the ACIS band.
For the measured $\Gamma$ and corresponding SED slopes, $B$ is insensitive to the upper energy, $E_M$, while the dependence on the lower energy, $E_m$, is more appreciable. 
For instance, the magnetic field for the near-half would change from $B_{\rm near}= 4.4\ k_m^{2/7}$ $\mu$G to $B_{\rm near}=15\ k_m^{2/7}$ $\mu$G if $E_m=0.1$ keV is replaced with $E_m=4.1\times10^{-9}$ keV (which corresponds to $\nu_m=E_m/h= 10 ^9$ Hz).  
For the far half, the $B$ values are very similar, $B_{\rm far }=4.1\ k_m^{2/7}$ $\mu$G and $B_{\rm far }=17\ k_m^{2/7}$ $\mu$G, for the same $E_m$ values, respectively. 
For equipartition,  the  magnetic field estimates imply  the tail energy densities  $w=w_B+w_e=(1.3-23)\times10^{-12}$ erg cm$^{-3}$, depending on the  choice of $E_m$.
Using the same method, we estimate the CN magnetic field to be $B_{\rm CN} \sim (16-28)\ k_m^{2/7}$ $\mu$G (depending on the value of $E_m$).
The Lorentz factor of electrons producing synchrotron photons with energy $E$ can be estimated as  $\gamma\simeq 1.4\times10^8 (E/1~{\rm keV})^{1/2}(B/10~{\rm \mu G})^{-1/2}$.  
For particles carried by the flow in the pulsar tail, the time they will radiate in X-rays is limited by the synchrotron lifetime, $\tau_{\rm syn}\simeq 5.1\times10^8 \gamma^{-1}B^{-2}$ s $ = 1100 (E/1~{\rm keV})^{-1/2}(B/10~{\rm \mu G})^{-3/2}$ years.
As no significant spectral cooling is seen in the B0355 tail, $\tau_{\rm syn}$ must be larger than the travel time $t_{\rm trav}\sim \mathcal{L}_{\rm tail}/u$, where $u$ is the bulk flow speed (assumed to be constant here for simplicity) and $\mathcal{L}_{\rm tail}=2.1$ pc is the length of the tail visible in X-rays.  
This requirement implies that the average flow speed $u\gtrsim\mathcal{L}_{\rm tail}/\tau_{\rm syn}=1800(E/1~{\rm keV})^{1/2}(B/10~{\rm \mu G})^{3/2}$ km s$^{-1}$, which significantly exceeds the  pulsar velocity.
Of course,  the initial flow velocity in the tail is likely to be much higher than the estimated average value (see simulations by B+05), and the flow should eventually slow down to be at rest with the ISM at large distances.

One can also use a simple 1D model of a collimated outflow cooling solely via synchrotron radiation (e.g., Chen et al.\ 2006) to estimate the flow speed. 
In a cylindrical tail where particles flow away from the pulsar with a constant velocity $u$ in a constant magnetic field $B$, the photon index at distance $z$ from the pulsar can be expressed as
\begin{multline}
\bar{\Gamma} = \frac{p+1}{2} \\ + \frac{p-1}{2} \frac{ ( \sqrt{\epsilon_m / \epsilon_l}-1)^{p-2} - ( \sqrt{ \epsilon_m / \text{min}(\epsilon_u,\epsilon_m)} -1)^{p-2} }{  ( \sqrt{\epsilon_m / \epsilon_l}-1)^{p-1} - ( \sqrt{ \epsilon_m / \text{min}(\epsilon_u,\epsilon_m)} -1)^{p-1}  }
\end{multline}
where $p=2\Gamma -1$ is the PL slope of the electron SED, which is assumed to produce a PL spectrum with the minimum and maximum energies, $\epsilon_l=0.5$ keV and $\epsilon_u=8$ keV, and 
\begin{equation}
\epsilon_m = 29 \left( \frac{z}{1~{\rm pc}} \right)^{-2} \left( \frac{u}{0.01 {\rm c}} \right)^2 \left(  \frac{B_\perp}{10~\rm{\mu G}}  \right)^{-3} \rm{keV}
\end{equation}
 is the photon energy corresponding to particles of maximum energy $E_m$ at distance $z$, with account for synchrotron losses.

\begin{figure}[h]
\epsscale{1.15}
\plotone{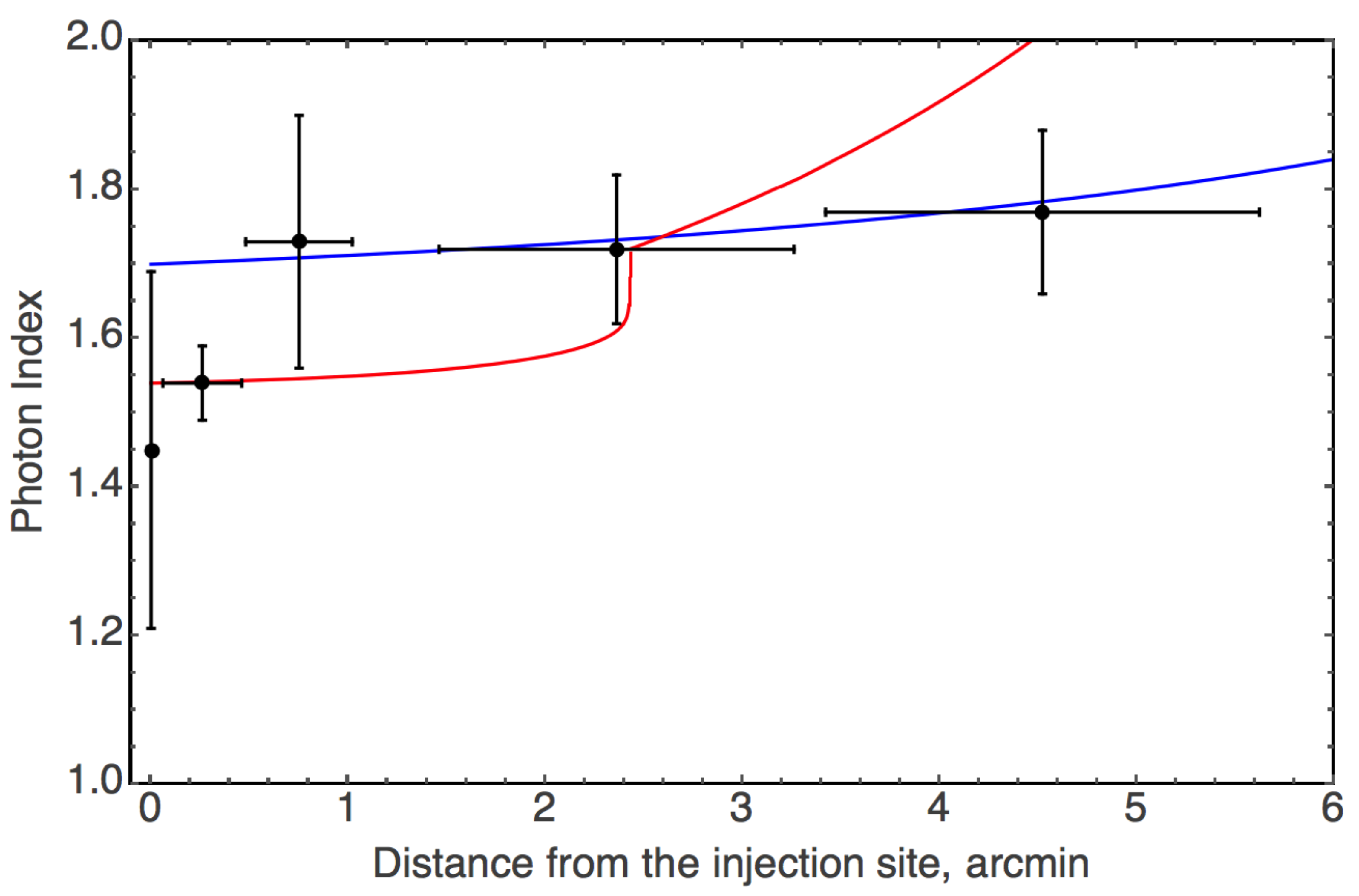}
\caption{Photon index as a function of distance from the pulsar.  Measured values are plotted as vertical black error bars ($1\sigma$) where the leftmost one represents the core (pulsar), and the subsequent ones are the CN, stem, near-half and far-half of the tail, respectively.  The horizontal bars indicate the distance over which the photon index was measured.  The red curve represents Eq.~(2) with the injection spectrum slope corresponding to that of CN  ($p=2.08$) and ``best-fit''  $(u/0.01~c)^2(B/10~\mu{\rm G})^{-3}= 0.15$.  The blue curve represents Eq.\ (2) with the injection spectrum slope corresponding to that of the stem ($p=2.46$) and $(u/0.01~c)^2(B/10~\mu{\rm G})^{-3}\sim 0.6$.  The break in the red curve occurs at the distance where $\epsilon_m = \epsilon_u$.}
\label{photon_index_distance}
\end{figure}

As one might expect, it is difficult to obtain a good correspondence between this simplistic model and the data, especially if the CN is included (red line in Figure \ref{photon_index_distance}). 
The model assumes constant tail cross section, magnetic field, and velocity, while all these parameters are likely to change substantially between the CN and the extended tail, and also within the tail.  
A seemingly satisfactory correspondence between the model and the data can be obtained if the injection is assumed to be happening further downstream (i.e. outside of CN) where the injection spectrum has a larger $p$ (corresponding to the somewhat softer $\Gamma$ than we measured).  
One can obtain reasonable fit as long as $(u/0.01~{\rm c})^2(B/10~{\rm\mu G})^{-3}\gtrsim 0.06$, corresponding to $u\gtrsim2400(B/10~\mu{\rm G})^{3/2}$ km s$^{-1}$, which is comparable to the simpler (monoenergetic) estimate given above.

For a cylindrical flow of radius $r_{\rm tail}$, the corresponding average energy injection rate into the tail can be estimated as $\dot{E}_{\rm tail} \sim \pi r_{\rm tail}^2 u (2w_B + \frac{4}{3}w_e)$.  
Assuming equipartition ($w_B = w_e$) in the B0355 tail, $\dot{E}_{\rm tail} \sim 1.1\times10^{34}\ {\rm ergs\ s}^{-1} (B/10\ \mu{\rm G})^2 (r_{\rm tail}/1.17\times10^{18}\ {\rm cm})^2 (u/2000\ {\rm km\ s}^{-1})$, implying that $\sim1/4$ of the pulsar's spin-down power ($\dot{E}=4.5\times10^{34}$) is transferred to the plasma flowing down the tail.

An alternative (to the high flow velocity) explanation of the slow cooling in the tail could be re-acceleration of the particles powered by turbulence in the tail plasma with the energy being taken from the magnetic field (e.g., via reconnection). 
However, the available X-ray data do not provide a way to discriminate between the two possibilities. 
The second scenario could be more favored in cases when the radio surface brightness of the tail increases with the distance from the pulsar (e.g., the tail of PSR J1509--5850; Ng et al.\ 2010, Klingler et al. 2016). 
So far, no radio emission has been detected from the B0355 tail.
Thus, sensitive lower frequency observations (e.g., IR and radio) are needed to determine a more accurate low energy break for the particle spectrum and a more accurate lower limit on the inferred magnetic field.

The above estimates neglect possible diffusion of the synchrotron emitting particles. 
Particle transport models with synchrotron radiation losses are widely used to simulate the spatial and spectral properties of PWNe such as the Crab Nebula, 3C 58, G21.5–--0.9 and others (see, e.g., Wilson 1972, and Tang \& Chevalier 2012; Porth et al.\ 2016). 
The diffusion coefficient is typically a free parameter in such models. 
The maximal spatial extension of the synchrotron tail in the one-dimensional diffusion model (assuming the lack of strong evolution of the magnetic field) $L_{\rm tail} \sim (2D \tau_{\rm syn})^{1/2}$. 
This allows us to constrain the diffusion coefficient $D$ of the synchrotron emitting electrons to be $D > 10^{27} (E / 1\ {\rm keV})^{1/2}(B/10\ \mu{\rm G})^{3/2}$  cm$^2$ s$^{-1}$.
Both Wilson (1972) and Tang \& Chevalier (2012) discussed simplified diffusion models in which $D$ is assumed to be energy-independent for synchrotron emitting electrons.
The constraints obtained above for the X-ray nebula created by PSR B0355+54 is consistent with the typical numbers used by Tang \& Chevalier (2012; see their Figure 9, where a much higher magnetic field was assumed to fit the radio through optical data of the Crab Nebula). 
Models of the plasma turbulence in relativistic outflows (Bykov \& Treumann 2011) typically predicted energy-dependent diffusion coefficients which are increasing with the particle energy. 
The constraint given above is actually close to the so-called Bohm diffusion coefficient, $D_B = c R_g/3$, where $R_g$ is the particle gyroradius. 
The value $D_B$ usually corresponds to the minimal possible value of the diffusion coefficient in the magnetic turbulence.
The above estimates may be even more appropriate for the whiskers (see Section 4.4) where they would lead to similar conclusions.

\subsection{Whiskers}
Outflows strongly misaligned with the pulsar velocity have been seen in three other PWNe created by supersonic pulsars:  the Guitar Nebula (PSR B2224+65; Hui \& Becker 2010), the Lighthouse PWN (IGR J11014--6103; Pavan et al.~2014, 2016) and the PWN of PSR J1509--5058 (Klingler et al.~2016).  
Bandiera (2008) suggested that such structures can occur when the stand-off distance at the apex of the bow shock, $r_s = (\dot{E}/4\pi c \mu m_p n v_p^2)^{1/2}$, becomes smaller than the electron gyration radius, $r_g=\gamma m_e c^2/eB_{\rm apex}$.
If true\footnote{This requirement may be violated if there is an ongoing reconnection between the internal PWN and external ISM magnetic fields, in which case particles with smaller $\gamma$ could leak as well (M.\ Lyutikov, private communication).}, this would imply that $B_{\rm apex}\lesssim  (4\pi c^5  m_e^2 \mu m_p n  \gamma^2 v_p^2)^{1/2}/e\dot{E}^{1/2}$ (here $B_{\rm apex}$ is the PWN magnetic field near the bow shock apex). 
In such a scenario, high energy particles can ``leak" into the ISM where they diffuse along the ambient magnetic field.  
Therefore, if the diffusion is fast, the orientation of the outflow reflects the geometry of the ambient magnetic field.  
As the whiskers do not appear to be bent back by the ram pressure, they are likely ambient ISM structures illuminated by pulsar-produced particles.

This scenario received support in the recent study of the Lighthouse PWN (Pavan et al.\ 2016) where evidence of draping of the ISM magnetic field lines around the bow shock head (Lyutikov 2006; Dursi \& Pfrommer 2008) is seen in the {\sl Chandra} ACIS image.
If the $r_s<r_g$ requirement is indeed a necessary condition for the leakage, it provides an additional link between pulsar and PWN properties, and decouples the ultra-relativistic electron energies from the often poorly constrained PWN magnetic field.
Indeed, the Lorentz factor of electrons escaping into the ISM is $\gamma\gtrsim 2\times10^{8}(E/1\,{\rm keV})^{1/2}(B_{\rm ISM}/5\,\mu{\rm G})^{-1/2}$ (here $B_{\rm ISM}$ is the ISM magnetic field), and hence one can constrain the magnetic field in the PWN head to be $B_{\rm apex}\lesssim7.6 (E/1\,{\rm keV})^{1/2}(B_{\rm ISM}/5\,\mu{\rm G})^{-1/2}\times (\sin i)^{-1}  n^{1/2}$~$\mu$G. 
This would imply that the magnetic field in the tail is likely close to the lower end of the range estimated in Section 4.3.1, and the magnetization should not be, in the Bandiera (2008) scenario, too far from the equipartition.

Consequently, the spectra of the misaligned outflows are expected to be relatively hard, as is observed in the Guitar ($\Gamma_{\rm B2224}=0.9^{+0.4}_{-0.2}$), the Lighthouse ($\Gamma_{\rm J11014}=1.6\pm0.2$), and the J1509--5850 ($\Gamma_{\rm J1509}=1.50\pm0.2$) spectra.  
In the case of B0355, the slope of the fitted PL is rather uncertain  ($\Gamma_{\rm B0355}=1.6\pm0.3$) but the best-fit value for the whiskers spectra  is smaller than that for any other region except the CN.  
In the other three known instances, the outflows display a puzzlingly strong asymmetry, being much more prominent on one side of the pulsar than on the other.  
The outflows seen in the Guitar, J1509--5850, and the Lighthouse PWNe also appear to be better collimated, which could be the result of stronger, more ordered ambient magnetic fields.  
The relative surface brightness of the B0355 whiskers (dimmer than the CN by a factor of $\sim$70) is comparable to the relative surface brightness of  J1509--5058's misaligned outflow (which is dimmer than the J1509--5058 CN by a factor of $\sim$50). 
More accurate spectral measurements for the extended tails and misaligned outflows should be possible with the {\sl Athena X-ray Observatory} (Kargaltsev et al.\ 2015).

\section{Summary}
We have presented a detailed analysis of the PWN created by the supersonic pulsar B0355+54.  
The deep \chan observations show the presence of a bright CN and a $7'$-long ($\sim2$ pc) tail directed opposite the pulsar's direction of proper motion.  

The CN displays a filled dome morphology, with the center being the brightest (in contrast to the Geminga and J1509--5850 CNe), a sharp trailing edge, and two faint protrusions (the split ``stem'' region) extending beyond this edge.  
We hypothesize that this is the result of our line-of-sight being directed along (or close to) the pulsar spin axis, with the jets then being bent back by the ISM ram pressure caused by the proper motion, superimposed on the CN, causing the center to appear brighter.  
With such interpretation, in ``hollow'' CNe morphologies (e.g., J1509--5850, Geminga), the brightened lateral regions could represent jets that are initially oriented at a larger angle with respect to the line of sight.

The emission from the pulsar and its immediate vicinity is softer, perhaps due to the presence of a thermal component, which could be attributed to emission from either the entire NS surface, implying $T\sim0.6$--$0.8$ MK, or from a $\sim250$ m hot spot with $T\sim1.8$ MK (for a BB model).

The tail shows no significant spectral softening up to 2 pc from the pulsar, suggesting a high bulk flow speed ($\gtrsim$ a few thousand km s$^{-1}$) or  in-situ particle reacceleration is occurring within the tail.

The deep observations also suggest the presence of two faint structures extending from the bow shock CN, roughly perpendicular to the pulsar's direction of motion.  
Similar outflows have been seen in other nebulae created by supersonic pulsars (the Guitar and Lighthouse PWNe), and are hypothesized to be the result of high energy electrons diffusing out of the bow shock region and interacting with the ambient ISM magnetic fields.

{\em Facilities:} \facility{{\sl CXO}}

\acknowledgements

We would like to thank Martin Weisskopf, Maxim Lyutikov, Giovanni Morlino, and Marina Romanova for the very helpful discussions.  
We are also grateful to the anonymous referee for the helpful suggestions and careful reading of the paper.
Support for this work was provided by the National Aeronautics and Space Administration through {\sl Chandra} Award Number G03-14082  issued by the {\sl Chandra} X-ray Observatory Center, which is operated by the Smithsonian Astrophysical Observatory for and on behalf of the National Aeronautics Space Administration under contract NAS8-03060.  
The work was also partly supported by NASA grant NNX08AD71G.  
A.~M.\ Bykov was supported by RSF grant 16-12-10225.

\end{document}